\documentstyle[aps,prb,epsfig,multicol,color]{revtex}

\begin{document}
\draft

\title{ Critical magnetic fluctuations induced
superconductivity   and residual density of states in $CeRhIn_5$
superconductor}

\author{Yunkyu Bang$^{1,2,3}$, I. Martin $^2$ and A.V. Balatsky $^2$}
\address{
$^{1}$ Department of Physics, Chonnam National University, Kwangju 500-757,
Korea}

\address{
$^2$ Theoretical Division Los Alamos National Laboratory, Los Alamos, New
Mexico \ 87545 }

\address{
$^{3}$ Center for Strongly Correlated Materials Research, Seoul National
University, Seoul 151-742, Korea}

\date{\today}
\maketitle

\begin{abstract}
We propose the multiband extension of the spin-fermion model to address the
superconducting d-wave pairing due to magnetic interaction  near critical
point. We solve the unrestricted  gap equation with a general d-wave symmetry
gap and find that divergent magnetic correlation length $\xi$ leads to the
very
unharmonic shape of the gap function with shallow gap regions near nodes.
These
regions are extremely sensitive to disorder. Small impurity concentration
induces substantial residual density of states. We argue that we can
understand
the large  $N_{res}(0) = \lim_{T\rightarrow 0} C_p(T)/T$ value and its
pressure
dependence of the recently discovered $CeRhIn_5$ superconductor under
pressure
within this approach.
\end{abstract}



\begin{multicols}{2}

Recent discovery of superconductivity in $CeMIn_5 (M=Co, Rh,
Ir)$\cite{Hegger,Petrovic} has spurred renewed interest of heavy fermion
systems and about the nature of its superconductivity. The rich phase diagram
of these compounds and the tunability by pressure and chemical substitution
of
the transition metal elements\cite{Pagliuso} provide  valuable information
about the competetion/interplay between the magnetism and superconductivity
and
perhaps a possible quantum criticality as a unifying origin of the phase
diagram of these compounds\cite{QCP}. While the complete understanding of the
phase diagram and the underlying mechanism for the magnetism and the
superconductivity is still lacking, there are  many details of thermodynamic
and transport properties of each phase in these materials which need to be
understood: large remnant $N_{res}(0) = \lim_{T\rightarrow 0} C_p(T)/T$
values
both in magnetic and in superconducting phases\cite{Fisher}, the weak first
order transition to superconductivity in $CeRhIn_5$\cite{Fisher}, a
unidentified new phase inside the mixed state\cite{Murphy}, a strong
deviation
of $\Delta C(T_c)/C(T_c)$ from BCS value and its pressure
dependence\cite{Fisher,Lengyel} etc.

Motivated by $CeRhIn_5$ experiment\cite{Fisher},
in this paper we examined  possible conditions for a d-wave
superconductors (1) to create a
substantial density of states (DOS) in superconducting phase
with a small impurity concentration, and (2) to have a large variation
of it as a function of pressure while keeping the constant
$T_c$\cite{Fisher}.

The $N_{res}(0)$ value at $16.5 kbar$ seen experimentally is almost  half of
the normal state  $N(0)$ \cite{Fisher}. From the previous
studies\cite{Hirschfeld}, it is required to have a large amount of impurities
in order to create  large $N_{res}(0)$ by magnetic/nonmagnetic isotropic
impurity in the superconducting state with lines of nodes\cite{Movshovich} (the
normal state scattering rate $\Gamma$ should be about $\Gamma/\Delta_0 \geq
0.5$ with Born scatterer and $\Gamma/\Delta_0 \geq 0.1$ with a unitary
scatterer\cite{Comment3}).
Even more intriguing is the pressure dependence of $N_{res}(0)$; it varies from
almost the half of the normal state $N(0)$ at 16.5$~kbar$  to almost zero value
at $21~kbar$ in one sample. Assuming  a simple form of d-wave paring $\Delta_0
(\cos(k_x)-\cos(k_y))$, such a large variation of $N_{res}(0)$ in the same
sample requires extremely sharp increase of $\Delta_0$ with increasing
pressure. This would be difficult to reconcile with almost constant $T_c$
values between $16 \sim 21 ~kbar$\cite{Fisher}.

To address these questions we need a microscopic model for the
superconducting state in $CeMIn_5$ materials. From the phase
diagram of these materials$^{1-5}$ and from the thermal
conductivity measurement indicating the unconventional pairing
with lines of nodes \cite{Movshovich}, we argue that it is highly
plausible that {\em the superconductivity is mediated by the
magnetic fluctuations}\cite{QCP,FirstPT}.
We propose a multiband generalization of the spin-fermion model\cite{Comment2}
where localized Ce spins $\vec{S}$ are interacting with the conduction
electrons (predominantly d band of $In$) via the Kondo exchange coupling $J$.
In mixed momentum and real space representation the Hamiltonian is written as
 \begin{equation}
 H = \sum_{{\bf k}, \alpha} c^{\dag}_\alpha({\bf
 k})\varepsilon({\bf k})c_\alpha({\bf
 k}) + \sum_{{\bf r},\alpha, \beta} J{\bf \vec{S}}({\bf r}) \cdot
c^{\dag}_\alpha({\bf r}){\bf \vec{\sigma}}_{\alpha \beta}c_\beta({\bf
 r}) + H_S
 \end{equation}
 where the first term is a kinetic energy
 and the second describes the Kondo exchange between
 Ce spins and conduction electron spin density. The last term
 represents an effective low energy Hamiltonian for the localized spins.
 The dynamics of the localized spins without long range  AFM order coupled
with  conduction electrons is well captured by the spin
 correlation function\cite{MMP,MBP,Chubukov},
$\chi({\bf q},\omega) = \delta_{ij}
 \langle S^i({\bf q},\omega)S^j({\bf q},-\omega)\rangle =
 \frac{V_0}{i\omega/\omega_0+ \xi^{-2} ({\bf q} -{\bf Q})^2 +1 }$,
where $\omega_0$ is  a spin relaxation energy scale, ${\bf Q}$ is the
2-dimensional antiferromagnetic  vector, and $\xi$ is the magnetic correlation
length.
The physics of this model for a one band case  has been investigated for a long
time \cite{MBP,Chubukov}. It is important to mention that the spin-fermion
model of our multiband case is different in the following aspects: (1) the
effective spin-fermion coupling should be much weaker \cite{Comment2} than that
of the one band model. One important consequence of it is that the relaxational
energy scale $\omega_0$ is much larger than that of the one band case
$\omega_{sf}$ \cite{MMP,Chubukov}; (2) more importantly, while the $\xi \gg a$
limit can not be reached in the self-consistent one band model without
deforming the FS topology of the conduction band \cite{Schmalian}, in the two
band model in conjunction with the weak coupling above mentioned, the $\xi \gg
a$ limit doesn't necessarily  modify the conduction band Fermi
surface\cite{Comment2};
(3) below $T_c$, however, the spin fluctuation damping $\omega_0$ is suppressed
for $\omega < 2\Delta_{sc}$, thus modifying the spin fluctuations spectrum.
This effect should be dealt with in the fully self-consistent Eliashberg
equation. In our weak coupling BCS approximation, this effect is ignored and
justified by the fact $\Delta_{sc} < \omega_0$ {\it a posteriori}.

The phase diagram for this material \cite{Fisher} suggests that $\xi$ becomes
gradually shorter away from the phase boundary ($P_c \sim 15 kbar$) toward the
higher pressure.
Then the basic feature of the magnetic fluctuations mediated
potential is that  near critical point with divergent $\xi$ $V({\bf q}) \equiv
\chi({\bf q}, \omega = 0)$ is sharply peaked around the antiferromagnetic wave
vector ${\bf \vec{q} \sim \vec{Q}}$ without much tail at other $\vec{q}$
values. With this potential the d-wave gap becomes mostly confined near the
antinodal points and the gap in the nodal regions becomes shallow\cite{Abanov}.
Upon decreasing the correlation length $\xi$, the potential has a longer tail
extending outside the ${\bf \vec{Q}}$ vector. As a result the gap function
approaches  a simple harmonic function $\sim (\cos(k_x)-\cos(k_y))$ with a
constant slope of the gap  over most of the Fermi surface (FS).
In this paper we show that the OP slope of the nodal region is very sensitive
to the shape of the pairing potential (see Fig.1(a)-(b)) and the shallow gap
near nodal points  leads to the enhanced sensitivity to disorder so that only
small amount of impurities is required to produce substantial residual DOS
$N_{res}(0)$. To capture this detailed shape of the d-wave OP, we solved the
unrestricted gap equation with only keeping the general d-wave symmetry. Then
we study the impurity effects using T-matrix approximation to calculate the
impurity-induced DOS $N_{res}(\omega)$ in the superconducting state.

{\em \bf Formalism.} For $CeRhIn_5$ we assume $\Delta_0 < \omega_0$. This
allows us to use a  weak coupling BCS gap equation of the spin-fermion model
for the d-wave pairing\cite{MBP}. For simplicity, we assume a circular FS in
two dimensions and integrate out the perpendicular component of momentum up to
the BCS cut-off energy $\omega_D$, which is naturally provided by $\omega_{0}$
in our model.
The effect of the impurity scattering is included within T-matrix
approxiamtion\cite{Hirschfeld}. For particle-hole symmetric case $T_3=0$, and
for d-wave OP with isotropic scattering $T_1=0$ (also without loss of
generality we can choose $T_2=0$ by U(1) symmetry). Then we need to calculate
only $T_0(\omega)$. Self energy is given $\Sigma_{0}=\Gamma T_{0}$, where
$\Gamma=n_i/\pi N_{0}$, $N_0$  the normal DOS at the Fermi energy, $n_i$ the
impurity concentration. Scattering strength parameter $c$ is related with the
s-wave phase shift $\delta$  as $c=\tan^{-1}(\delta)$. Now $T_0 (\omega_n)
=\frac{g_0 (\omega_n)}{[c^2-g_0 ^2 (\omega_n)]}$, where $g_0 (\omega_n) =
\frac{1}{\pi N_0}  \sum_k \frac{\tilde{i \omega}_n}{\tilde{\omega}_n^2 +
\epsilon_k^2 +\Delta(k)}$, $\tilde{\omega}_n=\omega_n+\Sigma_0$.
With this $T_0$ the following gap equation is solved self-consistently.
\begin{eqnarray}
\Delta(\phi) &=& - N_0 \int \frac{d \phi^{'}}{2 \pi} V(\phi-\phi^{'}) \cdot
F(\phi^{'}) \nonumber
\\ & & \cdot  T \sum_{\omega_n} \int^{\omega_D}_{-\omega_D} d \epsilon \frac{
\Delta(\phi^{'})}{\tilde{\omega}_n^2 + \epsilon_k^2 +\Delta(k)}.
\end{eqnarray}

Unlike the previous calculations of d-wave pairing, we assume no particular
functional form for $\Delta(\phi)$ except imposing $D_2$ symmetry; namely
$\Delta(n\pi/4)=0 (n=1,3,5,7) $, $\Delta(\phi)=\Delta(\phi\pm \pi)$, and
$\Delta(\phi)= -\Delta(\phi\pm \pi/2)$. Therefore the gap equation can
produce
the most general d-wave symmetry gap solution for a given pairing potential.
The  pairing potential is proportional to the static limit of $\chi({\bf
q},\omega=0)  \sim \frac{1}{(q-Q)^2+\xi^{-2}}$, which is parameterized as
follows.

\begin{equation}
V(\delta \phi)=V_d(b) \frac{b^2}{(\delta \phi \pm \pi/2)^2+b^2}.
\end{equation}

\noindent where the Lorenzian part is normalized and $V_{d}(b)$ determines the
total coupling strength. Apparently the parameter $b$ is proportional to
$\xi^{-1}$, normalized in the circular Fermi surface $(\xi \sim a \pi/ b$; $a$
is the lattice parameter). Both $\xi$ and $V_d(b)$ are functions of pressure.
We numerically determine $V_{d}(b)$   to make $T_c$ constant in accord with
$CeRhIn_5$ experiment\cite{Fisher}.
Finally, we introduce the FS weighting function $F(\phi)=\cos^{\beta}(2 \phi)$
to correct the artifact of the circular FS and to mimic the important aspect of
real FS topology\cite{beta}.

{\bf Results.} In all calculations we use $\omega_D=1$, which also serves the
unit energy. In Fig.1(a) we show the normalized pairing potentials
$V(\phi)/V_d(b)$ as a function of $b$ for illustration. In Fig.1(b) the
solutions of $\Delta(\phi)$ for the Born limit scatterer
($c=1$,$\Gamma=0.05$)
are shown for the potentials shown in Fig.1(a).

The self-consistently determined $T_0(\omega_n)$ is analytically continued to
real frequency using Pade approximant\cite{Vidberg} to calculate the
self-energy ($\Sigma_0(\omega+ i \eta)$). Then $N(0)= \frac{1}{\pi} \sum_k Im
G_0(\omega,k)$ is calculated. In Fig.2.a-b we plot $Im \Sigma_0 (\omega)$ for
both Born and unitary limits for  potentials with different $b \sim \xi^{-1}$
shown in Fig.1. There is no resonance at Fermi level for Born limit scatterer
as known, but the scattering rate $\gamma=Im \Sigma_0 (\omega=0$) increases as
the gap of nodal region becomes flatter, and for all cases for $|\omega| >
\Delta_0$ it approaches the normal scattering rate
$\Gamma_N=\Gamma/(c^2+1)=0.025$. As a result the residual DOS
$N_{res}(\omega=0)$ (see the Fig.3.a) sharply increases with flatter gap
region. On the other hand, for the case of unitary scatterer (Fig.2.b) there is
a resonance at the Fermi level for all cases, but the strength of this
scattering rate $\gamma$ has opposite trend in contrast to the Born limit; the
flatter the gap is, the smaller $\gamma$ is. This is because the
self-consistent equations are different for each case: $\gamma=\Gamma n_0 /c^2$
for Born limit and $\gamma=\Gamma /n_0$ for unitary limit
($n_0=<\frac{\gamma}{\sqrt{\gamma ^2 + \Delta(\phi)}}>$). Also for $|\omega|
> \Delta_0$ it approaches the normal scattering rate
$\Gamma_N=\Gamma=0.005$. Nevertheless for both cases even small
value of $\gamma$ is sufficient to create the substantial
$N_{res}(0)$ at Fermi level when the gap is shallow near nodes.
This is seen in Fig.3.a-b which plots the normalized DOS,
$N(\omega)/N_0$, for both Born and the unitary limits. The clearly
seen trend is  that the longer magnetic correlation $\xi$
produces a more residual DOS, $N(\omega)/N_0$ for a same amount of
impurities.

Fig.4(a) summarizes this trend, shows plots of $N(\omega=0)/N_0$
as a function of $b$ ($\sim P$) for both impurity cases. The
result shows that the Born limit scatterer has a stronger
dependence on pressure compared to the unitary scatterer. This is
because of the opposite trend of $\gamma$ in unitary scatterer due
to the resonant pole. In comparison to the experimental data of
$N_{res}(0)$\cite{Fisher}(shown in Fig.4(b)), the Born scatterer
fits  the data better.

In Fig.5 we show $T_c$ suppression as a function of impurity scattering
parameter $\Gamma$ for both Born ($c=1$) and unitary ($c=0$) limits for the
representative potentials ($b=0.1$). The impurity scattering parameters used in
our calculations gives  $(T_c(\Gamma)-T_{c0})/T_{c0}$ suppression of at most a
few $\%$ from Fig.5.(unitary limit $0.8 \%$, Born limit $4 \%$).
In passing  we note $\Gamma_{crit}/\Delta_0$ is about $50 \%$ larger than the
simple $\cos(2 \phi)$ type d-wave result\cite{Hirschfeld}  both for Born and
the unitary cases. This means that the unrestricted gap equation can find a
more optimized gap solution up to higher temperature compared to the fixed form
of $\Delta_0 \cos(2\phi)$ solution.

In summary, in this paper, we propose a multiband spin-fermion model as a
description of the pairing in $CeMIn_5$ materials. Motivated by the phase
diagram of these materials, we assume that the magnetic correlation length
$\xi$ decreases with pressure and the functional shape of the magnetic
fluctuations mediated pairing potential is changing. Using this potential we
show that the slope of the gap near nodes can be sharply changed. This strong
change of the slope can explain the pressure dependence of $N_{res}(0)/N_0$ as
well as the large value of it in $CeRhIn_5$ superconductor\cite{Fisher} close
to the quantum critical limit of $\xi \rightarrow \infty$ with a small amount
of impurity.

\begin{figure}
\epsfig{figure=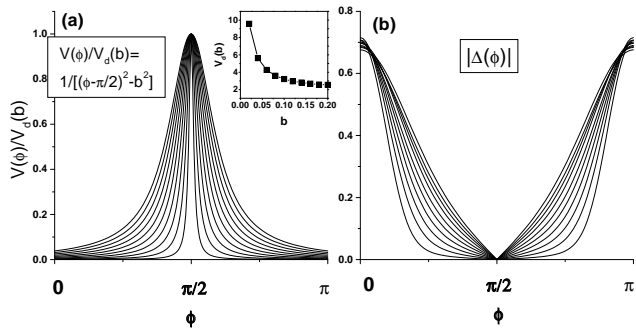,width=1.0\linewidth} \caption{(a) The normalized
pairing potential $V(\phi)/V_d(b)$ as a function of the exchange momentum
$\phi$, for different values of $b (\sim \xi^{-1})$. In increasing order of
potential width, $b$=0.02, 0.04, .... 0.2. $\phi=\pi/2$ is the AFM peak
momentum $\vec{Q}$. Inset is the $V_d(b)$ which is numerically determined to
make $T_c$ constant.  The trend is that the potential height is decreasing as
$\xi$ becomes shorter (or the pressure increases) and the width is increasing.
(b)The OP solutions $\Delta(\phi)$  for  pairing potentials shown in (a) with
impurities (Born limit $c=1$ and $\Gamma=0.05$) for all cases. In decreasing
order of potential width, $\Delta(\phi)$ becomes flatter near node. This trend
of the flatter gap near node approaching the magnetic phase (the longer $\xi$)
is also observed in the high $T_c$ superconductors\cite{Shen}. \label{fig1}}
\end{figure}

\begin{figure}
\epsfig{figure=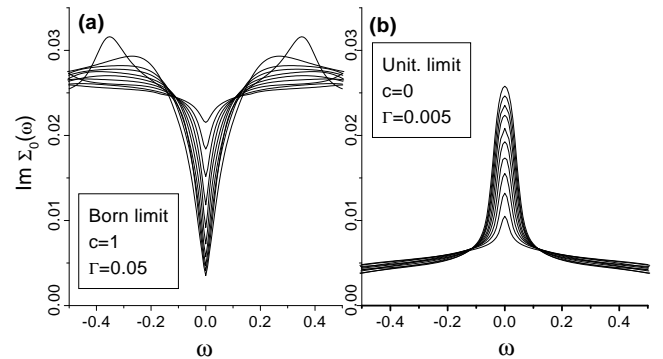,width=1.0\linewidth} \caption{(a) The imaginary part
of self-energy $Im \Sigma_{0}(\omega)$ for different pairing potentials  shown
in Fig.1. with Born limit scatterer ($c=1$ and $\Gamma=0.05$). With increasing
potential width, $\gamma=Im\Sigma_{0}(\omega=0)$ decreases.; (b) the same as
(a) with the unitary scatterer ($c=0$ and $\Gamma=0.005$). With increasing
potential width, $\gamma$ increases.
 \label{fig2}}
\end{figure}

\begin{figure}
\epsfig{figure=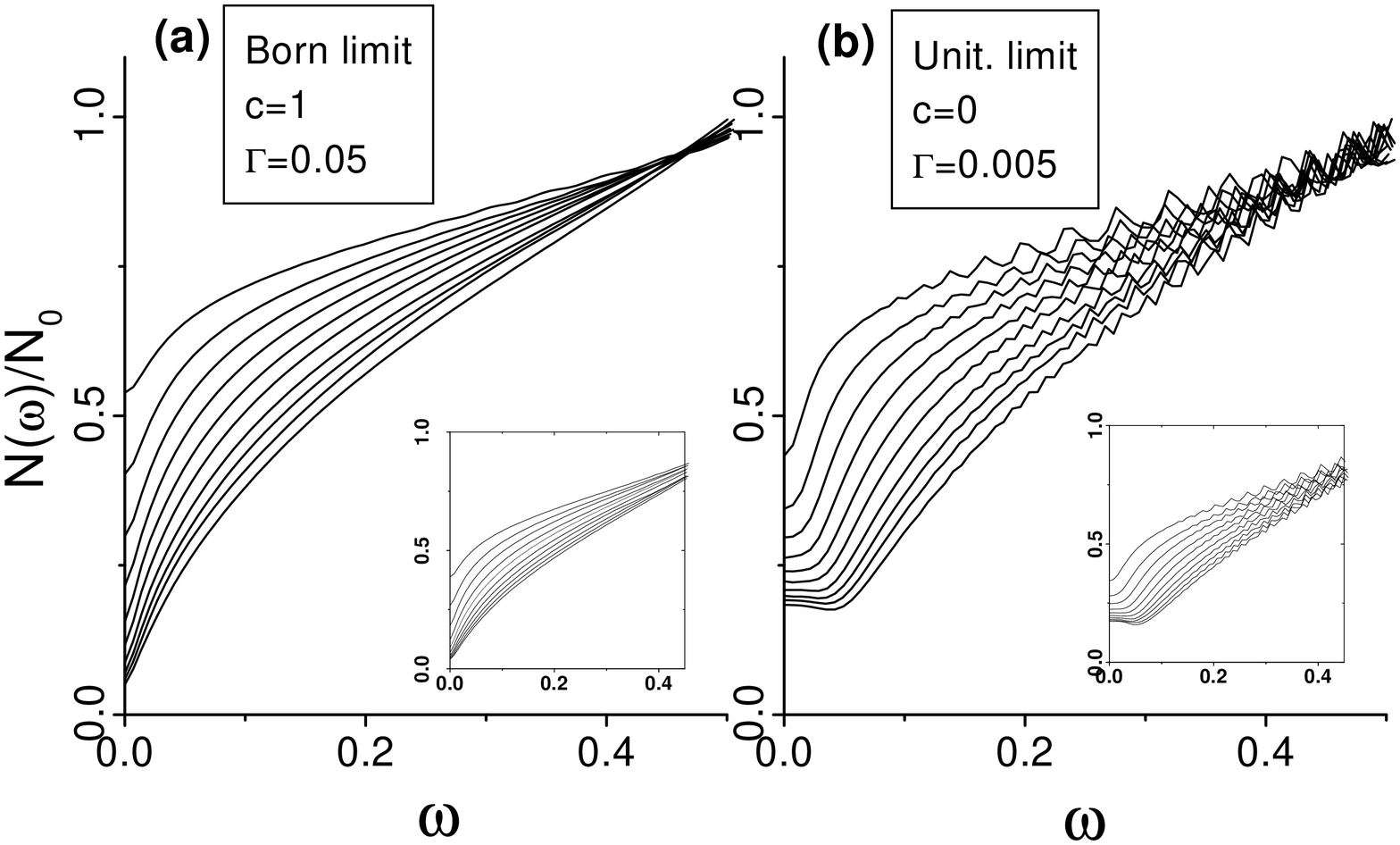,width=1.0\linewidth} \caption{(a) The normalized DOS
$N(\omega)/N_0$ for different  pairing potentials as shown in Fig.1. with Born
limit scatterer ($c=1$ and $\Gamma=0.05$). With increased   potential width
$N(\omega)/N_0$ decreases. Inset: with $\beta=4$ and $\Gamma=0.06$. (b) the
same as (a) with the unitary scatterer ($c=0$ and $\Gamma=0.005$). With
increased potential width $N(\omega)/N_0$ decreases. Inset: with $\beta=4$ and
$\Gamma=0.006$.
 \label{fig3}}
\end{figure}

\begin{figure}
\epsfig{figure=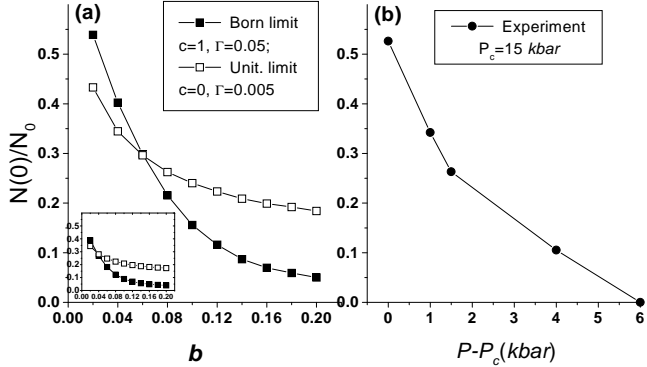,width=1.0\linewidth} \caption{(a) The normalized DOS
$N(\omega=0)/N_0$ as a function of $b$  ($\sim \xi^{-1} \sim P-P_c$) for Born
limit scatterer ($c=1$, $\Gamma=0.05$, solid square) and for the unitary
scatterer ($c=0$, $\Gamma=0.005$, open square). Inset is with $\beta=4$ ($c=1$,
$\Gamma=0.06$, solid square; $c=0$, $\Gamma=0.006$, open square); (b)
Experimental data from Ref[5].
 \label{fig4}}
\end{figure}

\begin{figure}
\epsfig{figure=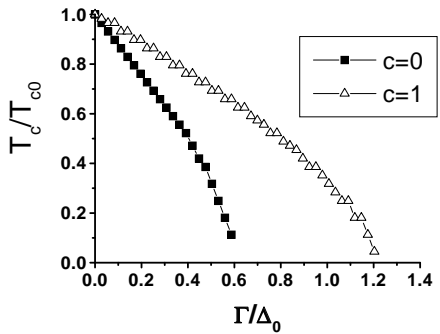,width=1.0\linewidth} \caption{$T_c/T_{c0}$ as a
function of $\Gamma/\Delta_0$ both for the Born ($c=1$) and  unitary limit
($c=0$) with a typical paring potential ($b=0.1$). \label{fig5}}
\end{figure}

We are grateful to R. Movshovich, J. Thompson, J. Sarrao, Ar. Abanov for
useful
discussions. This work was supported  by US DoE. Y.B. was partially supported
by the Korean Science and Engineering Foundation (KOSEF) through the Center
for
Strongly Correlated Materials Research (CSCMR) (2001) and through the Grant
No.
1999-2-114-005-5.

\end{multicols}


\begin{references}



\bibitem{Hegger}
H. Hegger et al., Phys. Rev. Lett. {\bf 84}, 4986 (2000).

\bibitem{Petrovic}
C. Petrovic et al., Europhys. Lett. {\bf 53}, 354 (2001); C. Petrovic et al.,
J. Phys. Condens. Matter {\bf 13}, L337 (2001).

\bibitem{Pagliuso}
P.G. Pagliuso et al.,  cond-mat/0107266.


\bibitem{QCP}
N.D. Mathur et al., Nature {\bf 394} 39 (1998); S.S. Saxena et al., {\it
ibid} {\bf 406} 587 (2000).

\bibitem{Fisher}
R. A. Fisher et al., cond-mat/0109221.

\bibitem{Murphy}
T. P. Murphy et al., cond-mat/0104179.

\bibitem{Lengyel}
E. Lengyel et al., unpublished.

\bibitem{Hirschfeld}
P. J. Hirschfeld and N. Goldenfeld, Phys. Rev. B {\bf 48}, 4219 (1993);
R.Fehrenbacher and M. Norman, Phys. Rev. B {\bf 50}, 3495 (1994).


\bibitem{Movshovich}
R. Movshovich et al.,  Phys. Rev. Lett. {\bf 86}, 5152 (2001).

\bibitem{Comment3}
If the data in Ref[5] were for a particular sample which happens to contain a
large amount of impurities, any cleaner sample should easily have twice as
large $T_c$. On the other hand the sample-to-sample variation of $T_c$ of
$CeRhIn_5$ is almost zero, so it means that the $T_c=2.3K $ sample is a
really
intrinsic one.


\bibitem{FirstPT}
The specific  heat data\cite{Fisher} suggests that the transition from the AFM
to SC is weakly first order, we think that this weak first order transition
doesn't change the nature of the AFM fluctuations drastically  in the SC phase.




\bibitem{Comment2}
Clear experimental evidence for the necessity of the two band model is that the
resistivity of $CeRhIn_5$ continues to be metallic and even decreased below
$T_{N}$ in the magnetic phase ($p < p_c$)\cite{Hegger}. It implies that the
long range magnetic order of $Ce$ spins doesn't induce the concomitant SDW gap
in the conduction band.

\bibitem{MMP}
A.J. Millis, H. Monien, and D. Pines,  Phys. Rev. B {\bf 42}, 167 (1990).


\bibitem{MBP}
P. Monthoux, A. V. Balatsky, and D. Pines, Phys. Rev. Lett. {\bf 67}, 3448
(1991);
Phys. Rev. B {\bf 46}, 14803 (1992);
K. Miyake, S. Schmitt-Rink, and C. M. Varma, Phys. Rev. B {\bf 34}, 6554
(1986).

\bibitem{Chubukov}
A. V. Chubukov, and D. Morr, Phys. Rep., {\bf 228}, 355 (1997) and references
therein.

\bibitem{Schmalian}
Jorg Schmalian, D. Pines, and B. Stojkovic, cond-mat/9804129.


\bibitem{Abanov}
Ar. Abanov, A. V. Chubukov, and A. M. Finkel'stein, Europhys. Lett. {\bf 54},
488 (2001).


\bibitem{beta}
Because of the circular FS, the pairing potential $V(\phi-\phi^{'})$ itself is
invariant with the simultaneous shift of $\phi$ and $\phi^{'}$, and cannot
distinguish the hot spots (parts of FS connected by the antiferromagnetic wave
vector $\pm \vec{Q}$) and cold spots on the FS. Therefore it is essential to
introduce the FS weight function  to avoid the artifact of the circular FS. In
this paper we use $F(\phi)=\cos^{\beta}(2 \phi)$ with $\beta=8$, but the
results with $\beta=4$ are not much different although it apparently produces
less flat OP near nodes.

\bibitem{Vidberg}
H.J Vidberg and J.W. Serene, J. of Low Temp. Phys. {\bf 29}, 179 (1977).

\bibitem{Shen}
Z.-X.Shen (private communication)




\end{references}
\end{document}